\documentclass[amsmath,amssymb,12pt,floatfix] {revtex4}
\usepackage[dvips]{graphicx}
\usepackage{bm}
\begin{document}
\title{Shock probes in a one-dimensional Katz-Lebowitz-Spohn model}
\author{Sakuntala Chatterjee (1,2) and Mustansir Barma (1)}
\affiliation{(1) Department of Theoretical Physics, \\
Tata Institute of Fundamental Research, \\
Homi Bhabha Road, Mumbai-400005, India, \\
(2) Institut f\"{u}r Festk\"{o}rperforschung, \\
Forschungzentrum J\"{u}lich, D-52425 J\"{u}lich, Germany}
\begin{abstract}
 We consider shock probes in a one-dimensional driven diffusive 
 medium with nearest neighbor Ising interaction (KLS model). Earlier studies
based on an approximate mapping of the present system to an effective
zero-range process concluded that the exponents characterising the decays of
several static and dynamical correlation functions of the probes depend
continuously on the strength of the Ising interaction. On the contrary, our 
numerical simulations indicate that over a substantial range of the
interaction strength, these exponents remain constant and their values are
the same as in the case of no interaction (when the medium executes an ASEP). 
We demonstrate this by numerical studies of several dynamical 
correlation functions for two probes and also for a macroscopic number of
probes.  Our 
results are consistent with the expectation that the short-ranged
correlations induced by the Ising interaction should not affect the large time
and large distance
properties of the system, implying that scaling forms remain the 
same as in the medium with no interactions present.
\end{abstract}
\maketitle
\section{Introduction}

Useful information about a complex system is often obtained by
introducing probe particles into it. After the probe particles have
come to a steady state with the system, their static and dynamic
behavior often reflect important characteristics of the system. For
instance, by monitoring the motion of probe particles, one can
understand visco-elastic properties of a cell~\cite{mason}, the
sol-gel transition in a polymer solution~\cite{nakanishi} or
correlations present in bacterial motion~\cite{shiva}.  In certain
cases, for example, in active micro-rheology, the probe particles are
subjected to external force fields. In~\cite{habdas} the forced
dynamics of a magnetic bead in a dense colloidal suspension, has been
used to study the colloidal glass transition.

In this paper we will consider one such example of nonequilibrium
(driven) probe particles which are introduced in a 
nonequilibrium medium, to study how the static and dynamic properties
of the probe particles are influenced by the surrounding medium, and 
also how the medium gets affected by the presence of the probe particles.

We study a particular simple one-dimensional lattice gas model first
introduced in \cite{kafri} to describe 
the motion of probe particles in a current-carrying medium.  The probe
particles are taken to exchange with particles and holes 
of the medium with equal rates but in opposite directions.  Because of
these dynamical rules, the probe particles 
tend to migrate towards the region of strong density variations (or
shocks) which may be present in the system.  Studying the dynamics of
these shock-tracking probe particles, one can therefore infer the motion of density
fluctuations in the medium.

In an earlier study~\cite{sc}, we have discussed the dynamical
properties of these probe particles in a nonequilibrium
current-carrying medium in which there is no interaction between
medium particles except hard-core exclusion. In this case, the medium
was described by an asymmetric simple exclusion process (ASEP) which
is the simplest lattice model of driven diffusive systems~\cite{liggett}. The
shock-tracking probe particles then reduce to second class
particles~\cite{kipnis}. Derrida {\sl et al.} have found the exact
stationary measure of the system~\cite{derrida}. Their studies on
static properties of the system show that when the number of second
class particles is finite, they form a bound state and the steady
state distribution function of the separation $r$ between a pair
decays as $r^{-\lambda}$ with $\lambda = 3/2$. A macroscopic number of
second class particles gives rise to a correlation length which
diverges (proportional to the square of the interprobe separation),
as the probe concentration goes to zero.

 We studied the dynamical properties of this system in
presence of a macroscopic number of probe particles, and found that
the dynamics is governed by a time-scale which marks the crossover
from single-probe behavior to many-probe behavior~\cite{sc}. This time-scale
shows a strong divergence (proportional to the cube of the
interprobe separation) in the limit of vanishingly small density of
the probe particles. This diverging time-scale is related to the
diverging correlation length present in the system~\cite{derrida}, and
enters the scaling descriptions of various dynamical correlation
functions of the probe particles ~\cite{sc}.
  
In the present paper, we present a detailed study of shock-tracking
probes in a driven system in which there is a short-ranged Ising interaction
between the particles of the medium. In the absence of any probes, such
a medium can be described by the $1$-d Katz-Lebowitz-Spohn (KLS)
model, whose steady state has 
an Ising measure~\cite{kls,luck}.
In~\cite{kafri} Kafri {\sl et al.} have reported 
that in presence of a macroscopic number of probe particles, the
system shows an interesting phase transition as the strength of the
Ising interaction is varied. Beyond a critical value of the
interaction strength and for sufficiently high density of the medium,
a macroscopic domain consisting of particles and holes (no probes) is
formed. A characterisation of this phase transition was attempted
using an approximate mapping to the zero-range process (ZRP) where the
probes are mapped onto ZRP sites and the particle-hole domain
preceding a probe is mapped onto the occupancy of that
site~\cite{kafri}. The current out of a particle-hole domain then
becomes the hopping rate out of a site in the ZRP.

A prediction of this approximate mapping is that exponents
characterizing the decays of several static and dynamic quantities
should depend continuously on the Ising interaction strength
$\epsilon$.  However, our numerical studies of these quantities seem
to indicate that the exponents are $\epsilon$-independent over a
substantial range of $\epsilon$.  This paper is concerned with a study
of the differences between our results and those based on the ZRP
picture.   

A possible simple rationalization of
our results is that the Ising interactions would be expected to give
rise to a finite correlation length $\xi_{\rm Ising}$, whose 
value may be renormalized in the presence of probes, but is still
expected to be finite. Then, on length
scales $r \gg \xi_{\rm Ising}$, the system should behave essentially
like the non-interacting $(\epsilon = 0)$ system.  Our results are
indeed consistent with such a scenario, as we find 
$\epsilon$-independent behaviour asymptotically (for
large $r$ and $t$), even though there is sometimes an $\epsilon$
dependence for smaller $r$ and $t$.

Below we describe in brief the quantities we studied and our
results. 
\begin{enumerate}
\item[{(1)}] {\it Distribution function of the size of the
particle-hole domains}: The mapping to the ZRP predicts that in the
  disordered phase this
distribution function should be an exponential times a power law, with
a power which is a continuous function of the Ising interaction
strength.  However, we observe that the domain size distribution shows a power
law exponent which does not vary with 
$\epsilon$ but remains constant at its value for $\epsilon = 0$. To
understand this discrepancy, we are  
led to check the assumptions that have been made in the approximate
KLS-ZRP mapping. We find that the assumption of statistical
independence of the domains remains valid, and further verify
that accounting for the finite size correction to domain currents is
not the reason 
behind the discrepancy. However, as we discuss in section $3$, the
movement of probes in a KLS medium is non-Markovian and the ZRP
mapping does not capture this aspect of probe dynamics.  

\item[{(2)}] {\it Dynamics of two probe particles}: This was
studied in~\cite{levine,rakos} where it was reported that a bound state
forms between the probe pair such that the distribution of separation
decays as a power law with an exponent $b(\epsilon)$ that varies
continuously with the strength of the Ising interaction
$\epsilon$. Starting from a configuration where the two probes were
nearest neighbors, a scaling form was proposed to describe the
temporal evolution of their separation. The authors 
have tried to verify this scaling form by measuring the cumulative distribution
and the mean value of the separation between the probes as a function
of time. They reported that in conformity with their scaling
hypothesis, the time-dependent cumulative distribution function $\tilde{P}(r,t)$ for
different values of $t$ undergoes a scaling collapse when rescaled by
$t^{[b(\epsilon)-1]/z}$ and plotted against $rt^{-1/z}$, where $z$ is the
dynamical exponent that takes the value $3/2$. The average separation between the 
two probes is reported to grow with time as a power law with an exponent
$[2-b(\epsilon)]/z$, which is consistent with their scaling form.

On the contrary, we find that although for an initial time range, the
average distance does show $\epsilon$-dependent growth, 
for larger times, it crosses over to another growth regime
where the exponent takes a value which is close to the one expected
for $\epsilon =0$, {\sl i.e.} with no Ising interaction. Our numerical results
also show that for larger times, the scaling collapse of $\tilde{P}(r,t)$
fails. We have verified that the scaling collapse can be retrieved 
by rescaling with $t^{1/3}$ (instead of $t^{[b(\epsilon)-1]/z}$), as in the
case of $\epsilon=0$.

\item[{(3)}] {\it Dynamical properties with a macroscopic number
of probes}: The dynamical correlation functions in this case are found
to follow the same scaling description as with $\epsilon =0$, with a
crossover time-scale which separates a single-probe regime at short
times from a long-time regime characterized by collective behaviour of
the probes. Moreover, the crossover time-scale shows a similar divergence
in the limit of 
vanishingly low concentration of the probe particles. In other words,
our studies indicate that even in the presence of a nearest neighbor
Ising interaction in the medium, the large time and large distance
properties of the system do not change.
\end{enumerate}
   
In the following section, we describe the lattice model on which we
have performed Monte Carlo simulation and briefly summarise our earlier
results for the non-interacting $(\epsilon = 0)$ medium. In section
$3$ we discuss the static properties of this model where we recall the
approximate mapping to the zero-range process (ZRP) introduced
in~\cite{kafri} and discuss the validity of various assumptions that
went into this mapping. In section $4$ we discuss the dynamical
properties of the system in presence of a finite number of probes and
also for a finite density of the probes.

\section{Model and Earlier Results $(\epsilon =0$)}

The model is defined on a one dimensional periodic lattice each site of which
may either be empty or may contain a particle of the medium or a probe. We
use the symbol `$+$' to denote a particle, `$-$' to denote a hole and `$0$' to
denote a probe. The exchange rules are as follows.
\begin{eqnarray}
\nonumber
+- &\stackrel{\mbox {\normalsize $1-\Delta V$}} \longrightarrow & -+ \\
\label{eq:klsrate}
+0  &\stackrel{\mbox {\normalsize 1}} \longrightarrow & 0+ \\
\nonumber
0-  &\stackrel{\mbox {\normalsize 1}} \longrightarrow & -0
\end{eqnarray}
Here $\Delta V$ is the change in the nearest neighbour Ising interaction
potential
\begin{equation}
V= - \frac{\epsilon}{4} \sum_i s_i s_{i+1} 
\label{eq:Hkls}
\end{equation}
where $s_i = 0, \pm 1$, according as the site $i$ contains a probe, a
particle or a hole, respectively. Throughout we consider equal densities of
particles and holes in the medium, {\sl i.e.} $\rho_0 = 1-2\rho$ where 
$\rho$ and $\rho_0$ denote densities of particles and probes, respectively.
The coupling parameter $\epsilon$ may vary
in the range $[-1,1]$. In this paper, we will only consider $\epsilon > 0$.   

In the absence of any probes, the system reduces to a $1$-d KLS model with an
Ising measure in the steady state. This gives rise to an $\epsilon$-dependent 
 correlation length $\xi_{\rm Ising}$ in the system. For $\epsilon <1$, this
correlation length remains finite and hence the large distance properties of
the system can be expected to remain  unaffected by the interaction.

When probes are present, as seen from the last two exchange rules in 
Eq. \ref{eq:klsrate}, a probe
exchanges with particles and holes of the medium in opposite
directions. This implies that a probe would tend to be located in a
position where there is an 
excess of holes to its left and an excess of particles to its right. In other
words, there would be a strong density variation or `shock' around a probe.
This is the reason we call them `shock-tracking probes' (STPs).

In the absence of any interaction, one has $\epsilon =0$ and in this case,
 a particle in the medium executes a totally
asymmetric exclusion process (TASEP) with an effective hole density
$(1-\rho)$ and it exchanges with a hole and a probe
in the same way; similarly, a hole in the medium also executes a TASEP [in the
opposite direction and with an effective particle density $(\rho+\rho_0)$] and
it exchanges with a particle and
a probe in the same way. In other words, for $\epsilon =0$ a probe behaves
like a particle for an adjacent hole, and like a hole for an adjacent particle.
 Such probes are known
as `second class particles'~\cite{kipnis}.

Derrida {\sl et al.} ~\cite{derrida} have found the exact steady state
measure of this system 
of second class particles in an ASEP by using the matrix method. In presence
of more than one second class particle, the steady
state factorises about any second class particle, which implies factorisation
in terms of the one component system about the shock position. When
there is a single 
second  class particle present in the system, the mean density profile
around it decays as a
power law with an exponent $1/2$. In the presence of two (or a finite
number of) 
second class particles, the medium induces an attraction between them and they
form a weakly bound state where the distance $r$ between two successive second
class particles follows a power law distribution $P(r) \sim r^{-3/2}$. When
the number of second class particles is macroscopic, the density profile at a
distance $r$ from any given probe takes the form 
\begin{equation}
\rho (r) \sim \frac{1}{\sqrt{r}} \exp \left ( - r/\xi \right ) + \rho
\label{eq:xi1}
\end{equation}  
where the correlation length $\xi$ diverges in the low concentration limit of
the probes~\cite{derrida}:
\begin{equation}
\xi \approx 4 \rho (1-\rho)/\rho_0 ^2 \;\;\;\;\;\;\;\; {\mbox {as $\rho_0 \rightarrow
0$}}
\label{eq:xi2}
\end{equation}

We monitored several quantities to study the dynamical properties of
systems with macroscopic number of probes.  We find 
a diverging time-scale which marks the crossover between
single-probe behavior and many-probe behavior. In Section IV, we will
discuss the behaviour of these quantities, when $\epsilon$ is nonzero
in the KLS model.  

The variance of the displacement of the tagged probes is defined as 
\begin{equation}
C_0 (t) = \langle \left ( Y_k (t) - Y_k (0) - \langle Y_k (t) - Y_k (0)
\rangle \right ) ^2 \rangle 
\label{eq:deftag}
\end{equation} 
where $Y_k(t)$ is the position of the $k$-th probe at time $t$. Ferrari and
Fontes~\cite{ferrari} had earlier calculated the asymptotic ($t
\rightarrow \infty$) 
behavior of $C_0(t)$  
 and shown that $C_0(t) \approx Dt$ with diffusion constant $D = \left [\rho(1-\rho)
+ (\rho + \rho_0 )(1-\rho - \rho_0 ) \right ] /\rho_0$. For small times, in
the limit of low concentration of the probe particles, one would expect each
probe to behave as an individual non-interacting particle subject only to the
fluctuations of the medium. The variance of the displacement of a single probe
is found analytically to grow as $t^{4/3}$~\cite{evans,kirone}. In the limit
of small but finite concentration of the probe particles $C_0(t)$ shows a
single-particle (super-diffusive) behavior at small time and diffusive
behavior at asymptotically large times. One would therefore expect a crossover
between these two regimes that would occur at a time-scale $\tau$ which is a
function of $\rho_0$. The natural expectation would be $\tau \sim \xi^z$ where
$\xi$ is the correlation length as defined
in Eq \ref{eq:xi1}. Substituting the value of the dynamical exponent $z=3/2$
and using Eq. \ref{eq:xi2} one obtains 
\begin{equation}
\tau \sim \rho_0 ^{-3}
\label{eq:tau}
\end{equation}  
in the limit of small $\rho_0$. This leads us to propose the following scaling
form  for $C_0(t)$
\begin{equation}
C_0(t)  \sim t^{4/3} F \left ( \frac{t}{\tau}  \right ). 
\label{eq:stptag}
\end{equation}
This form is valid in the scaling
limit of large $t$ and large crossover time-scale
$\tau$ ({\sl i.e.} $\rho_0 \rightarrow 0$).
Here $F(y)$ is a scaling function which approaches a constant as 
$y \rightarrow 0$.
For $y \gg 1$, we must have  $F(y) \sim y^{-1/3}$, in order to reproduce
$C_0(t) \approx Dt$. 
We have verified the above scaling form by Monte Carlo simulation~\cite{sc}.

The same crossover time-scale $\tau$ is found to be present in other dynamical
correlation functions as well. To track the dissipation of the density
pattern of the second class particles, we considered the quantity
\begin{equation}
B_0(t)=\overline{\left ( Y_k(t)-Y_k(0)-\overline{(Y_k(t)-Y_k(0))}\right )^2}
\label{eq:defvan}
\end{equation}
where the overhead bar denotes averaging over different evolution histories,
starting from a {\it fixed} initial configuration drawn from
the steady state ensemble (see ~\cite{shamik,beijeren} and
also~\cite{sc} for a discussion on why this special averaging process
is useful in 
tracking dissipation). Our scaling analysis leads to the following scaling
form:
\begin{equation}
 B_0(t) \sim t^{4/3} G\left ( \frac{t}{\tau} \right ). 
\label{eq:van}
\end{equation}
where $\tau$ is the same crossover time-scale as in Eq.\ref{eq:tau} and
 $G(y)$ is a scaling function which 
approaches a constant as $y \rightarrow 0$, while for $y \gg 1$, one
expects $G(y) \sim y^{-2/3}$. Our numerical results are consistent with this
scaling form~\cite{sc}.

Finally consider the quantity
\begin{equation}
\Delta (t) = \langle \left ( R(t)-R(0)\right ) ^2 \rangle
\label{eq:defdelta}
\end{equation}
which measures how the separation between two successive probes fluctuates in
time. Here, $R(t)$ is the separation between the $k$-th and $(k+1)$-th pair at
time  $t$. Our studies show that $\Delta (t)$ has the following scaling form
\begin{equation}
\Delta (t) \sim t \;H \left ( \frac{t}{\tau } \right )
\end{equation}
where the scaling function $H(y)$ approaches a constant as $y
\rightarrow 0$ and for $y \gg 1$ one must have $H(y) \sim 1/y$.

To summarise, for $\epsilon=0$ we find that several dynamical correlation functions
of the probe particles are governed by a single crossover time-scale $\tau$ which
diverges as $\rho_0 ^{-3}$ for low concentration of the probes. In the
remaining portion of the paper, we will consider static and dynamical
properties for $\epsilon > 0$ and examine how different they are from the
non-interacting case.

\section{Static Properties of KLS Model with Probes}

Kafri {\sl et al.} reported 
that the KLS model with macroscopic number of
probes shows phase separation transition for $\epsilon > 0.8$ as the
density $\rho$ is increased above a critical value $\rho_c$
~\cite{kafri}. They concluded that in the phase
separated state, a macroscopic domain, composed of particles and holes of the
medium, coexists with another phase which consists of small domains of
particles and holes, separated by the probes. They explained this phase
transition by attempting to approximately map
the system onto a zero-range process.

To describe the mapping, we first define a domain as an uninterrupted sequence
of particles and holes, bounded by probes from both ends. The current $J_n$
out of a domain of length $n$ can then be determined by studying a KLS model
in an open chain with boundary rates of injection and extraction equal to 
the rate at which the particles and holes of the domain would exchange with
the probes at the domain boundaries. According to Eq. \ref{eq:klsrate} this
rate is unity. The current $J_n$ can be calculated exactly for an open KLS
chain and for large $n$ it has the form 
\begin{equation}
J_n= J_{\infty} \left ( 1 + \frac{b(\epsilon)}{n} \right )
\label{eq:jn}
\end{equation}
where the coefficient $b$ has the following dependence on $\epsilon$
\begin{equation}
b(\epsilon) = \frac{3}{2} \frac{(2+\epsilon)v+2\epsilon}{2(v+\epsilon)}, 
\;\;\;\;\;\;\;\;\; v=\sqrt{\frac{1+\epsilon}{1-\epsilon}} +1. 
\label{eq:epb}
\end{equation}
The study of Kafri {\sl et al.} indicates that $b$ plays an important role in
characterising the phase separation transition in the model.

\begin{figure}[!htbp]
\begin{center}
\includegraphics[scale=0.8,angle=0]{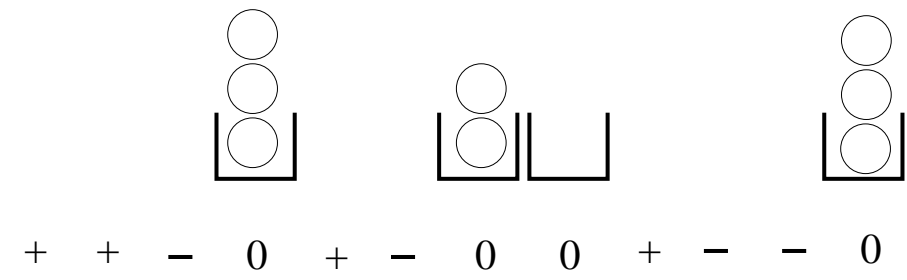}
\caption{\it A typical configuration of the KLS model with probes and its
corresponding configuration in ZRP.}
\label{fig:zrp}
\end{center}
\end{figure}

The present system is mapped onto a zero-range process (ZRP) as follows: the
$i$-th probe is defined as the $i$-th site of ZRP and the length of the domain
to the left of the $i$-th probe is taken to be the occupancy $n(i)$ of the
$i$-th site of ZRP. We illustrate this in fig \ref{fig:zrp}. The density
in the ZRP is related to the KLS model density $\rho$ as $\rho_{ZRP} =2\rho/
(1 - 2\rho) $.

The hopping rate out of the
$i$-th site in the ZRP is taken to be the domain current $J_{n(i)}$
given in Eq. 
\ref{eq:jn}. For such a ZRP, the condition for condensation to 
take place is 
$b>2$ and $\rho_{ZRP} $ larger than a certain critical density $ \rho_c$.  In the condensed 
phase, the occupancy at a single site becomes
macroscopically large, while the remaining sites have an average occupancy
$\rho_c$ ~\cite{martinzrp}. For $\rho_{ZRP} < \rho_c$, the number of
particles present on a site follows the distribution function
\begin{equation}
P(n) \sim \frac{1}{n^b}\; exp(-n/\xi_{\rm ZRP})
\label{eq:zrpcl}
\end{equation}
where the correlation length $\xi_{\rm ZRP}$ diverges as $\rho_{ZRP} \rightarrow
\rho_c$.  For $\rho_{ZRP} = \rho_c$, we have $P(n) \sim \displaystyle\frac{1}
{ n^b}$, while for $\rho_{ZRP} > \rho_c$, a similar power law decay
describes the distribution at all sites except for the single condensate
site.

The approximate ZRP correspondence implies that in the   
KLS chain with probes, 
for large enough $\rho$ and for $\epsilon > 0.8$ (as follows from
Eq. \ref{eq:epb}), there 
should be a macroscopic domain present in the system which
is composed of particles and holes (no probes). 
The rest of the system should consist of small probe clusters, interrupted by  
the domains (of particles and hole) with size distribution given by 
Eq. \ref{eq:zrpcl}.

From numerical simulations for $\epsilon < 0.8$, it was found that a
very large domain may 
exist for large $\rho$ ~\cite{kafri}. Our numerical simulations confirm this.
In~\cite{kafri,torok} it has been argued that this is not true phase
separation, but rather a consequence of the fact that the correlation
length in this case has a large (but finite) value.

\begin{figure}[!htbp]
\begin{center}
\includegraphics[scale=1.0,angle=0]{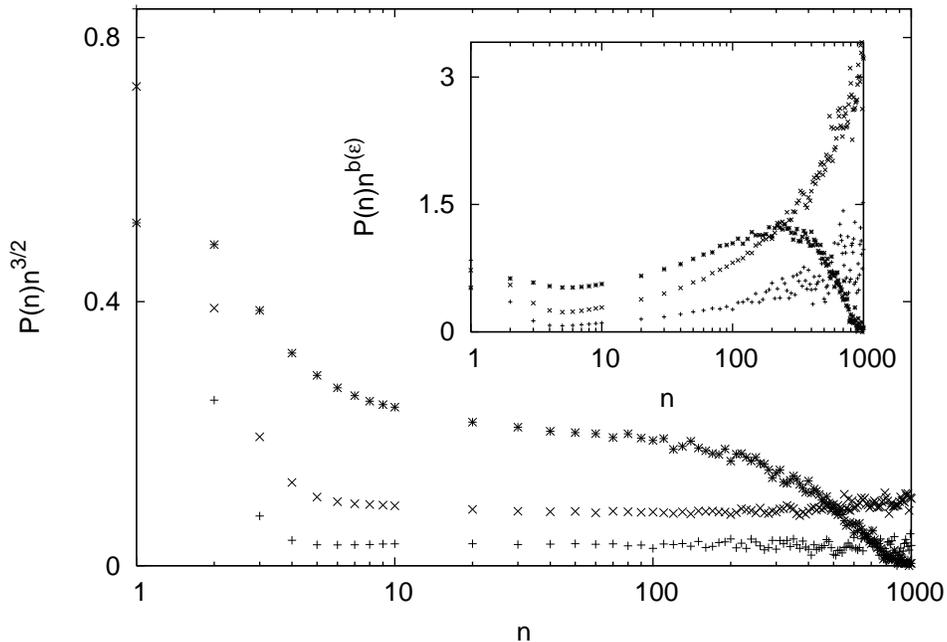}
\caption{\it Domain size distribution $P(n)$ scaled by $n^{3/2}$ shows a flat
stretch over a substantial range of $n$ for different values of $\epsilon$ and
$\rho$. For comparison with the ZRP prediction, we have scaled $P(n)$ by
$n^{b(\epsilon)}$ in the inset. In both these plots, the symbol $\bm +$
corresponds to $\epsilon=0.9,\rho=0.464$, symbol $\bm \times$ corresponds to 
$\epsilon=0.8, \rho = 0.375$ and symbol $\bm \ast$ corresponds to $\epsilon=0.6,
\rho=0.375$. In the last case, $\epsilon$ is substantially smaller than the critical value and
$\xi_{ZRP}$ is shorter. This explains the observed deviation from the power
law behavior for large $n$.}
\label{fig:domaindist}
\end{center}
\end{figure}

According to the above correspondence with the ZRP, it is expected that close
to the critical point, the
domain size distribution for $n \ll \xi_{\rm ZRP}$ should follow a power law 
with exponent 
$b(\epsilon)$ which should increase monotonically with $\epsilon$. 
However, our numerical simulations for various values 
of $\epsilon$ and $\rho$ [see fig \ref{fig:domaindist}] 
show that the power law exponent seems throughout to be close to
$3/2$ (which is the value of $b$ at $\epsilon =0$),
independent of the value of $\epsilon$. 

This points to a contradiction with the correspondence with the ZRP, 
and leads us to examine the
assumptions that go into the KLS-ZRP  mapping.

\begin{figure}[!htbp]
\begin{center}
\includegraphics[scale=1.0,angle=0]{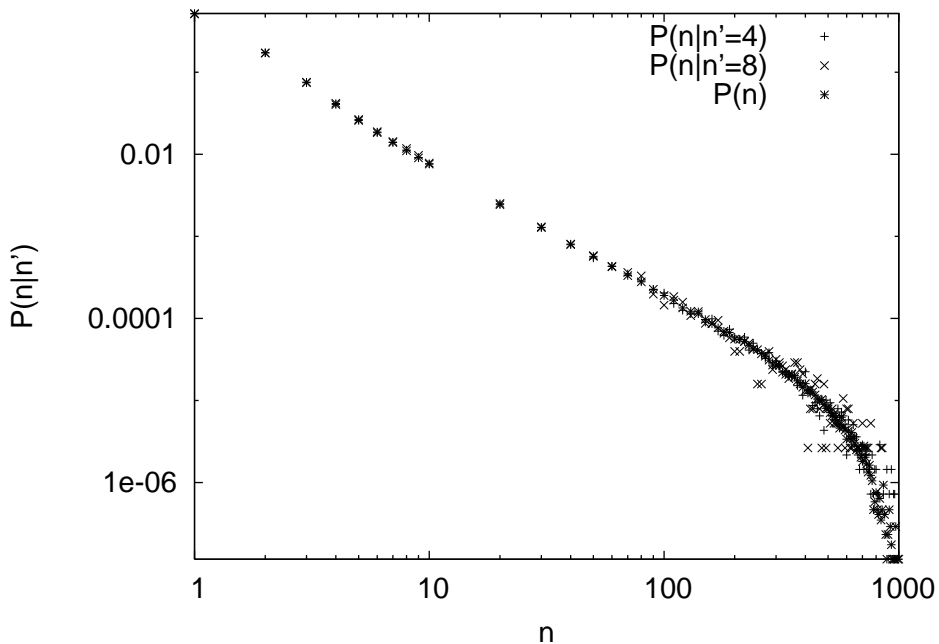}
\caption{\it The conditional distribution of domain size $P(n|n')$ as a
function of $n$ for $n'=4,8$. For comparison $P(n)$ is also shown. $P(n|n')$
is seen to match with $P(n)$ which shows the domains are independently
distributed. We have used $L=2048$, $\epsilon=0.6$ and $\rho=0.375$.}
\label{fig:condclust}
\end{center}
\end{figure}

{\it Independence of Domains}: A crucial property of the ZRP is
that the occupancies at the sites are uncorrelated. In our present
model of the 
KLS chain with probes, this would imply that the domains between the probes
should be
independently distributed. We have verified this assumption by measuring
the conditional probability $P(n|n')$ that the size of a particular 
domain is of length $n$ given that its neighboring domain is of length $n'$.
We find that $P(n|n')$ does not depend on $n'$ and is same as $P(n)$
consistent with neighboring domains being distributed
independently. Our data is 
presented in fig \ref{fig:condclust}.

{\it Finite Size Correction to Domain Current}: Apart from the
independence of domains, another requirement for the ZRP mapping to hold
is that the current out of a domain of size $n$ should be the same as
the current in an 
isolated open KLS chain and is given by Eq. \ref{eq:jn}.
Evans {\sl et al.} have given evidence for this by 
numerically measuring the actual current out of a domain and comparing 
with the exact calculation for an open chain KLS
model~\cite{mohanty}. Good agreement was 
found for large $n$.

To take into account the finite size corrections for moderate values
of $n$, we simulated a ZRP where the hopping rate out of a site is
read off directly from the actual $J_n$ vs $n$ data, obtained from
numerical simulation of the KLS model with probes.  The mass
distribution for this ZRP is found to have the same form as in
Eq. \ref{eq:zrpcl} with the exponent $b$ given by Eq.  \ref{eq:epb},
as expected. We conclude that the finite size correction to $J_n$ is
not the reason for the discrepancy shown in fig \ref{fig:domaindist}.

{\it Non-Markovian Movement of the Probes}: There is however,
one aspect of the KLS model with probes that is not captured in the
corresponding ZRP. Since a probe exchanges with the particles and holes of the
medium in opposite directions, as shown in Eq. \ref{eq:klsrate}, once a probe
moves in one particular direction, it cannot move in the opposite
direction at
the very next time-step. For example, suppose a probe moves to the left by
exchanging with a particle in the medium.
Immediately after this exchange the probe has the
particle as its right neighbor. Clearly, the probe cannot take a step to the
right as long as that particle stays there. In other words, the probes have a
finite memory which makes their movement non-Markovian. 
 In terms of the ZRP this would mean that
once a site has emitted a particle to its right neighbor,
it has to wait for some time till it can receive a particle from its right
neighbor.
This waiting time should depend on the form of the density profile in a
domain. Note that in this non-Markovian ZRP, apart from $J_{\infty}$ and
$b(\epsilon)$, there are other parameters that are associated with the exact
form of the waiting time. As a result, the phase-diagram becomes complicated
and to specify the criterion of a phase transition a much more detailed
analysis is required. This might shed some light on the  observed discrepancy
about domain size distribution.

\section{Dynamics of Probes in the  KLS Model}
\label{sec:klsdyn}

\subsection{Two Probes}

The properties of two STPs in the KLS chain were first studied    
in~\cite{levine} by Levine {\it et. al.}, who argued that the
time-evolution of the separation between the probe pair is governed by
a Master equation. Their analysis indicates  
that the medium induces an attraction among the probe particles and they form
a bound state. The steady state distribution of the distance between two
probes takes the form $P(r) \sim r^{-b}$ where $b$ is a function of
$\epsilon$ given by Eq. \ref{eq:epb}. For $\epsilon=0$ one retrieves $P(r)
\sim r^{-3/2}$ as found  in~\cite{derrida}.

Rakos {\sl et al.} have shown that the random force between
the probe pair is sensitive to the noise correlations present in the
medium~\cite{rakos}.  When
the probe particles are embedded in a KLS ring, such that the random force
that drives the probe particles is fully generated by the current fluctuations
of the driven medium, the probes inherit the dynamical exponent of the medium,
which is $3/2$. On the other hand, if the random force has a part that is
temporally uncorrelated, the resulting motion is described by a dynamical
exponent $z=2$.

To study the dynamics of the system, the distance between the two
probes was monitored, starting from the initial configuration in which the
two probes were side by side. The approach to the steady state was modelled
by the scaling ansatz
\begin{equation}
P(r,t) \sim r^{-b} f(r/t^{1/z})
\label{eq:levscale}
\end{equation}
where $P(r,t)$ is the probability that starting as nearest neighbors,  
the two probes are at a distance $r$
apart at time $t$. In the range $1<b<2$ this would imply that the average distance
between the two probes grows as 
\begin{equation}
\langle r(t) \rangle \sim t^{(2-b)/z}.
\label{eq:levav}
\end{equation}
Since $b$ is an increasing function of the Ising interaction
$\epsilon$, this would predict a slower growth law of $\langle r
\rangle$ with $t$, as $\epsilon$ increases.

The cumulative distribution function $\tilde{P} (r,t)$ is defined as the
probability that starting from a nearest neighbor position, the separation
between the two probes at time $t$ is larger than $r$. From Eq.
$\ref{eq:levscale}$ it follows that 
\begin{equation}
\tilde{P} (r,t) \sim t^{(1-b)/z} Y(r/t^{1/z})
\label{eq:cum}
\end{equation}  
which means that $\tilde{P} (r,t)t^{(b-1)/z}$, plotted against $r/t^{1/z}$
should show a scaling collapse for various values of $t$.

In~\cite{levine,rakos} the time evolution of the average distance
between the two 
probes was monitored numerically. Starting from a randomly disordered
configuration, with the restriction that the two probes are placed on nearest
neighbor sites, the system was evolved for  a time $t_{equil}$
in an attempt to let it reach an equilibrium state. The time evolution
during the equilibration process 
followed the exchange rules shown in Eq. \ref{eq:klsrate} with the
important modification
that the two probes were constrained to remain nearest neighbors {\sl i.e.}
they
hopped together as if glued together. At the end of this
equilibration, the medium 
is assumed to be locally in steady state, in the
vicinity of the probes, up to a distance of the order $t_{equil}^{2/3}$. At
this point, defined as $t=0$, 
the restriction for the relative position of the probes was
released and the distance between them monitored. The distance between
the probes was then assumed  
to follow the scaling form in Eq. \ref{eq:levscale} for $t \ll t_{equil}$ when
the two probes move within an equilibrated region. In this time regime,
it was numerically verified that the growth of $\langle r(t) \rangle $
is consistent with Eq.
\ref{eq:levav}~\cite{levine,rakos}.

\begin{figure}[!htbp]
\begin{center}
\includegraphics[scale=1.0,angle=0]{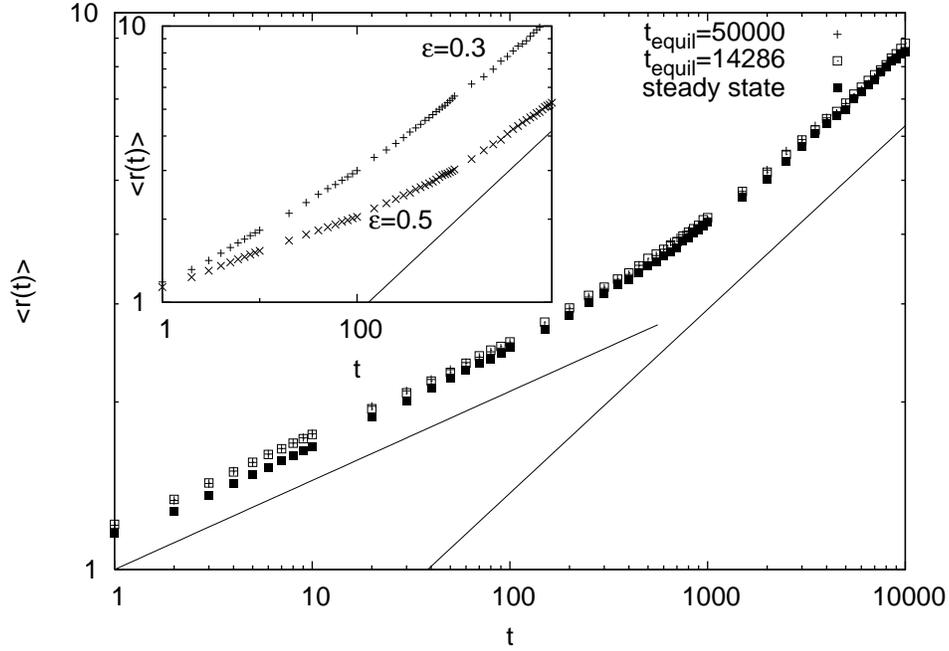}
\caption{\it Average distance $\langle r(t) \rangle $ between the probe pair
as a function of time. $\langle r(t) \rangle$ shows two different power law 
growths as time changes. The reference lines show that the 
growth exponent is $(2-b)/z$ at short times and changes to $1/3$ at large
times. The curves for partially equilibrated initial conditions
(using the method of Rakos {\sl et al.}) with different
values of $t_{equil}$ coincide for small $t$. We have also measured 
$\langle r(t) \rangle $ starting from steady state initial condition.   
The partially equilibrated data
and steady state data coincide for large $t$. We have used $\epsilon =0.4$ and
$L=1000$. Inset shows the steady state data for $\epsilon=0.3,0.5$ with
$L=4096$  and the reference line with exponent $1/3$.}
\label{fig:2zrav}
\end{center}
\end{figure}

Note that the scaling form in Eq. \ref{eq:levscale} is expected to be valid in
steady state. Therefore, to verify this scaling form, we followed the
following procedure.  Allow the
system to reach steady state by evolving it without any restriction
on the relative separation of the two probes.  Then
wait till the probes come to a nearest neighbor position with respect to each
other and define $t=0$ at this point.  Our
data shows that $\langle r(t)\rangle $ follows Eq. \ref{eq:levav} only for an
initial time-regime, after which the growth exponent changes to
$\simeq 1/3$ which
is close to the value of the growth exponent at $\epsilon=0$. We present our
data in fig \ref{fig:2zrav}.

We have also measured $\langle r(t) \rangle $ following the procedure
of~\cite{levine,rakos}. We investigated the effect of  
different values of $t_{equil}$ and found the same
behavior as described in the previous paragraph. Moreover, fig \ref{fig:2zrav}
shows that the curves for this
partially equilibrated initial condition, coincide with that of the steady
state
initial condition (explained in the previous paragraph), for large time.    

We conclude that in steady state,  $\langle r(t) \rangle$ does not follow Eq.
 \ref{eq:levav} with an $\epsilon$-dependent $b$ all the way, but shows a 
crossover at large time to the behavior $t^{1/3}$, which is the behavior  
obtained for $\epsilon =0$.

\begin{figure}[!htbp]
\begin{center}
\includegraphics[scale=1.2,angle=0]{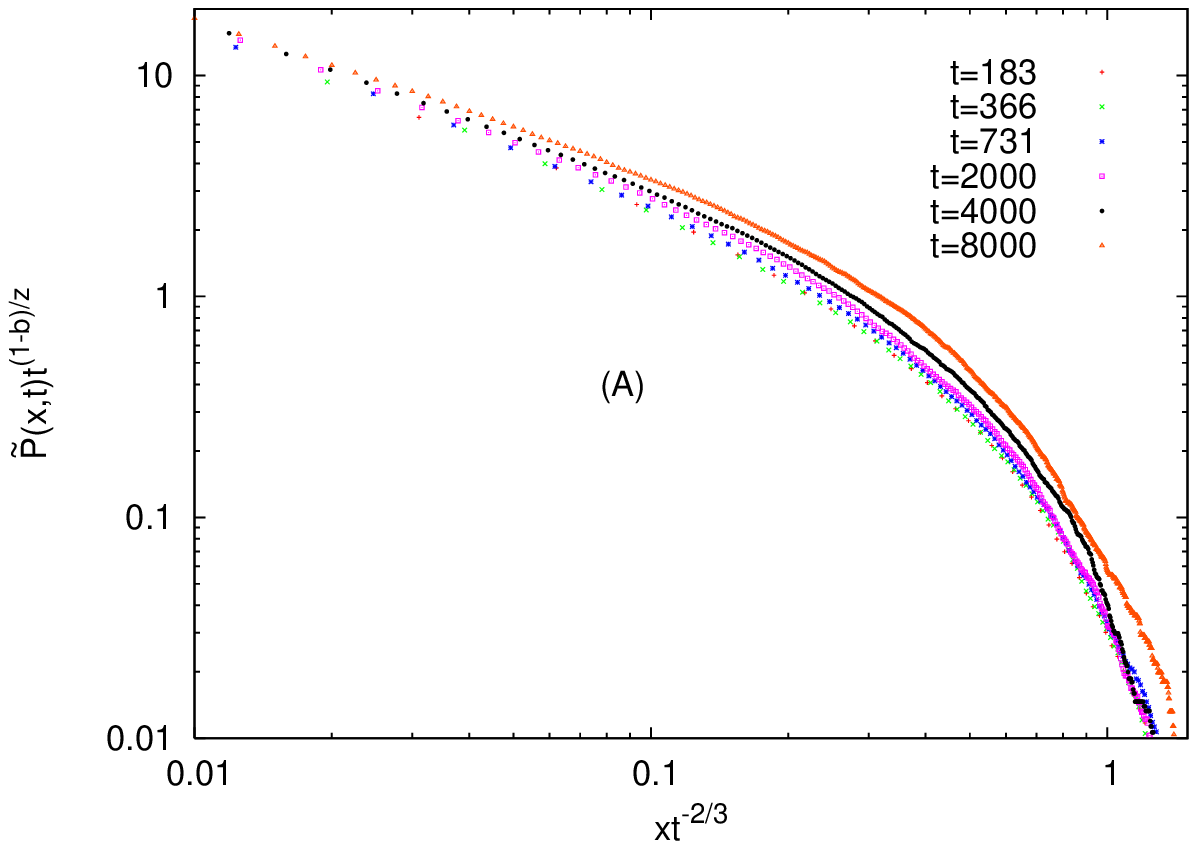}
\includegraphics[scale=1.2,angle=0]{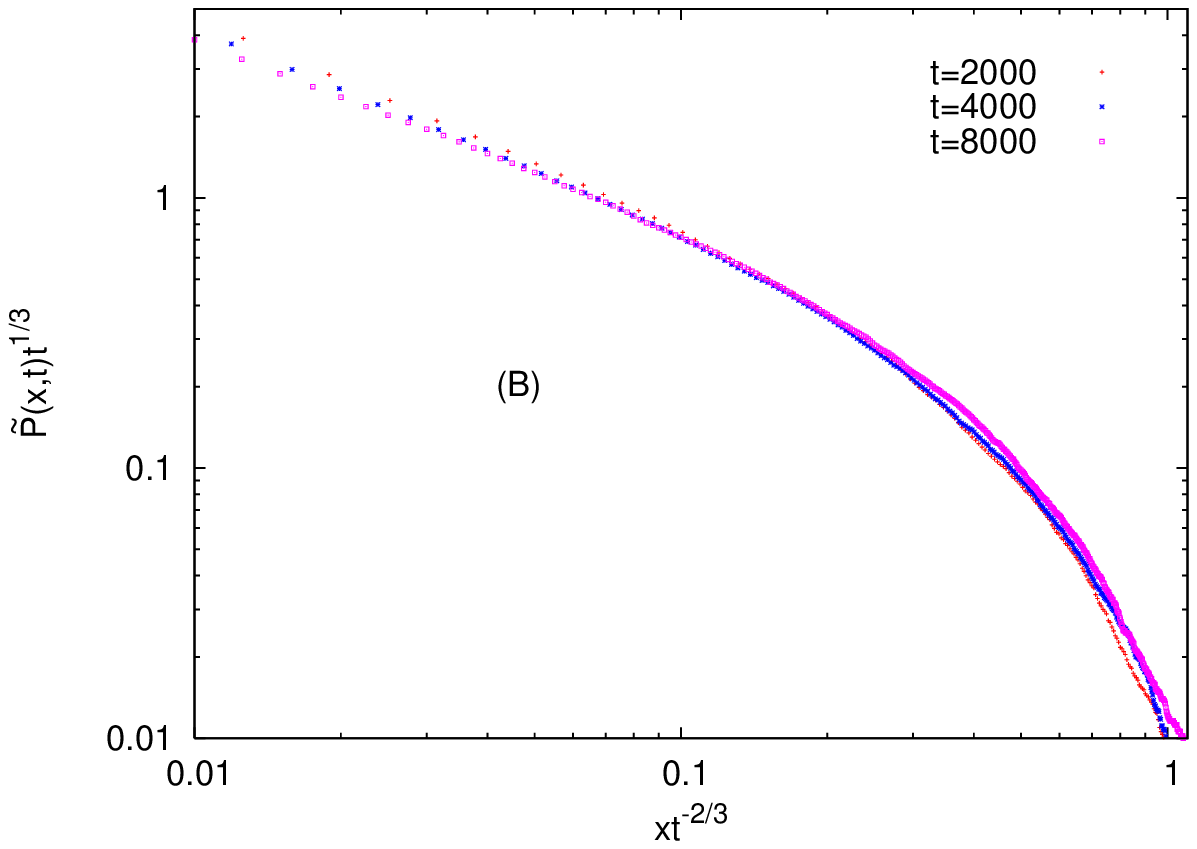}
\caption{\it Color online: Panel $(A)$ shows the lack of scaling collpse for the cumulative
distribution function $\tilde{P}(x,t)$ for larger values of $t$. As seen from
the label of $y$ axis,
when $\tilde{P}(x,t)$ is rescaled by an $\epsilon$-dependent prefactor, the
scaling collapse works for smaller $t$, but fails for larger $t$.This is in
contradiction with the scaling form described in Eq. \ref{eq:cum}. Instead,
$\tilde{P}(x,t)$ is seen to follow a
scaling form $\tilde{P}(x,t)\sim t^{-1/3}X(x/t^{2/3})$, for larger values of
$t$, as shown in panel $(B)$. We have used $L=4096$ and $\epsilon=0.4$.}
\label{fig:distb}
\end{center}
\end{figure}

In~\cite{rakos} it was also reported that the cumulative distribution function
$\tilde{P} (r,t)$ shows a scaling form as in Eq. \ref{eq:cum}. Starting from
an initial configuration with the two probes next to each other (as discussed
above), $\tilde{P} (r,t)$ was numerically measured for a range of values of
$t$ and it was concluded that within that range, $\tilde{P} (r,t) t^{2(b-1)/3}$
shows a scaling collapse for different values of $t$, as plotted against
$r/t^{3/2}$. However, our numerical results indicate that this scaling collapse 
fails for larger $t$ values (see fig \ref{fig:distb}A). 
Instead, an $\epsilon$-independent scaling form,
 more specifically, the scaling form expected for $\epsilon=0$, seems to hold.
We show this by plotting $\tilde{P} (r,t) t^{1/3}$ against $r/t^{3/2}$ and fig
\ref{fig:distb}B shows the scaling collapse for larger $t$ values.

In our simulation, we could not go to very large times as the finite size
effects would become strong. In fig \ref{fig:distb} 
we have presented our data for the largest
system size $(L=4096)$ we could access. However, the crossover time is
much smaller and no finite size effects are observed for this time-range.

Our studies therefore show that the large time dynamics of the two probes is not
affected by the presence of an interaction in the medium as reported
in~\cite{levine,rakos}, rather it resembles the case of non-interating
medium. As discussed in Section 1, one possible rationalization is that the  
Ising measure of a KLS model (without probes) induces a finite 
$\epsilon$-dependent correlation length  in the medium. In the presence of
 probes, the value of this
correlation length may change, but it is expected to be finite still.  
As long as the displacement of the
probes is less than this correlation length, the effect of varying $\epsilon$
may be felt. But asymptotically, when the typical probe
separation has exceeded the 
Ising correlation length, it is plausible that they behave as if in a medium
with no interactions, i.e. $\epsilon =0$ 
\footnote{Throughout this paper, we have considered probe particles embedded in a KLS
ring and found a large time crossover in the probe dynamics as discussed
above. Instead of a KLS model, when the probe particles are embedded in a medium,
such that certain movements of the probes become temporally uncorrelated
(see~\cite{rakos} for details), then no such crossover effects have been
observed. In this case of ``mixed dynamics" of the probe particles, an
$\epsilon$-dependent scaling form seems to remain valid, within the range of
time we have investigated.}.

\subsection{Macroscopic Number of Probes}

We now take up the study of a system with a macroscopic number of
probes.  We find that the dynamics of
the STPs is governed by a diverging time-scale $\tau$, as in  
the non-interacting case $\epsilon =0$. For $t \ll \tau$, an STP senses
the fluctuations solely due to the KLS chain. But a KLS chain is known
to have an 
Ising measure which means that if $\epsilon$ is not too close to unity, only
short-ranged correlations are present in the medium. Let $\tau_0$ be
the time required for a probe particle to move a distance of order
$\xi_{\rm Ising}$.  Then 
for $\tau_0 \ll t \ll \tau$, the dynamics of the probes in a KLS chain
should be similar to those in 
an ASEP (where no correlation is present in the medium) i.e. as that
of the second class particles discussed 
in~\cite{sc}. The dependence of the crossover time $\tau$ on the probe
density is discussed below.

Let $r_i$ be the separation between the $i$-th and $(i+1)$-th probe and $R_m$
be the distance between the first and the $(m+1)$-th probe, {\sl i.e.} $R_m =
\sum_{i=1}^m r_i$. Let $r_i$ follow the distribution $P(r_i) \sim r_i
^{-\lambda}$. Assuming independence, the quantity $R_m$ which is the sum
of $m$ such random 
variables should follow a L\'{e}vy distribution with a norming constant $\sim
m^{1/(\lambda-1)}$, so long as $R_m$ is less than the correlation length
$\xi$. 
In other words, the length $R_m$ of a 
segment which contains $m$ probes scales as $m^{1/(\lambda-1)}$. This  
is valid up to $R_m \sim \xi$ but fails as $R_m$ increases 
beyond that. $\xi$ is the same correlation length that appears in
Eq. \ref{eq:xi1} for the non-interacting case. 
Let $m_>$ be the number of STPs in a segment of length $\xi$.
Then $m_> \sim \xi ^{\lambda-1}$.
Hence in a system of length $L$, the total number
of probes $N_0$ can be written as
$N_0 = \left ( L/ \xi \right ) \xi ^{\lambda-1}$,
which implies that the correlation length $\xi \sim \rho_0^{-1/(2-\lambda)}$
and
hence $\tau \sim \xi ^{z_0} \sim \rho_0^{-z_0/(2-\lambda)}$, where $z_0$ is
the
dynamical critical exponent of the system.

\begin{figure}[!htbp]
\begin{center}
\includegraphics[scale=1.0,angle=0]{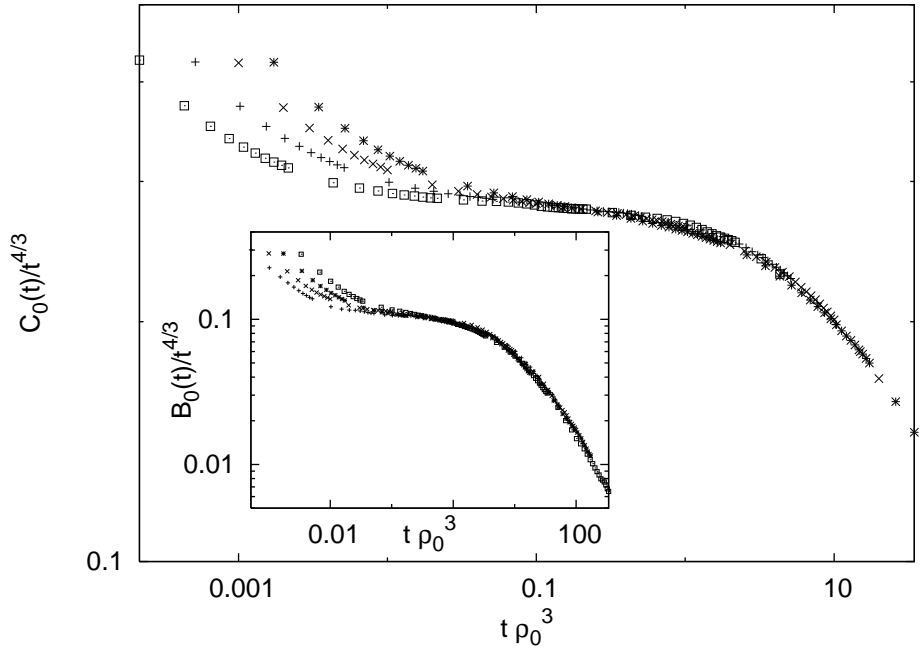}
\caption{\it Scaling collapse for $C_0(t)$ for $\epsilon =0.2$ and $\rho_0
=0.06,0.08,0.1,0.12$. Inset shows scaling collapse for $B(t)$ with 
$\epsilon =0.2$ and $\rho_0 =0.08,0.1,0.12,0.15$. We have used $L=16384$. }
\label{fig:eptag}
\end{center}
\end{figure}

\begin{figure}[!htbp]
\begin{center}
\includegraphics[scale=1.0,angle=0]{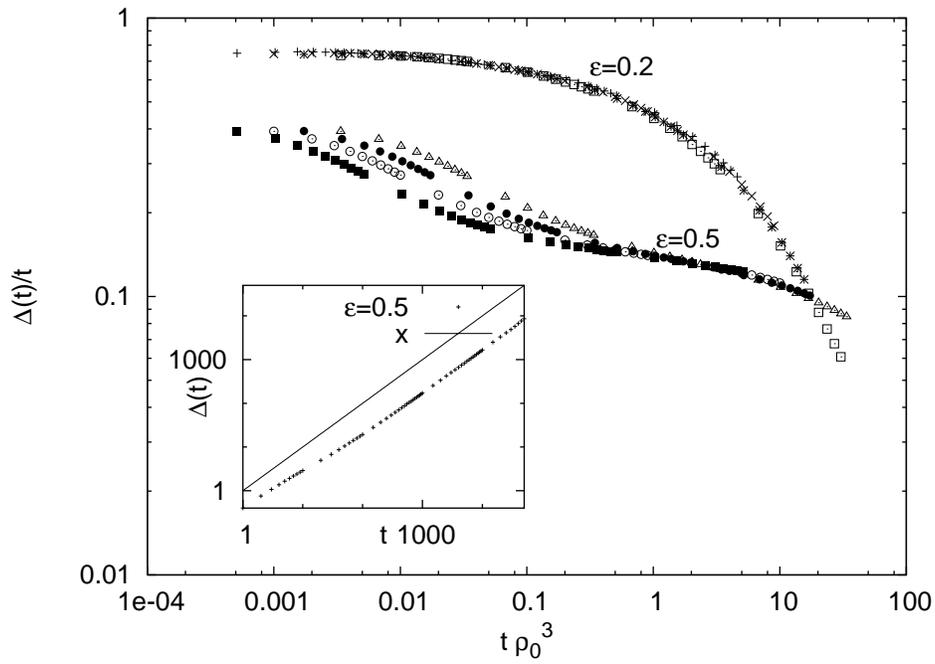}
\caption{\it Scaling collapse for $\Delta (t)$ for finite $\epsilon$ values.
We have used $\rho_0=0.08,0.1,0.12,0.15$ and $L=16384$. The inset shows the
linear growth of $\Delta (t)$ for $L=16384$ in presence of two probes.}
\label{fig:epdelta}
\end{center}
\end{figure}

We have monitored the dynamical 
correlation functions $C_0(t)$, $B_0(t)$ and $\Delta (t)$, as defined in Eq.
\ref{eq:deftag}, \ref{eq:defvan} and \ref{eq:defdelta}, respectively. Our
numerical simulations indicate that these quantities follow the same scaling
form as in the non-interacting case $\epsilon =0 $~\cite{sc}.
 Moreover they continue to
show crossover at a time-scale $\tau \sim \rho_0 ^{-3}$, very similar
to the $\epsilon =0 $ case. In fig \ref{fig:eptag} we show the scaling
collapse for $C_0(t)$ and $B_0(t)$.  We present our data for $\Delta (t)$ in fig 
\ref{fig:epdelta}.

In the case of two probes, one might expect $\Delta (t)$ would  
show the same scaling behavior as the second moment of the distribution
$P(r,t)$ in Eq. \ref{eq:levscale}, {\sl i.e.} $\Delta (t)$ should grow with
time as $t^{(3-b)/z}$. But our numerical simulations show
that irrespective of the value of $\epsilon$, $\Delta (t)$ always grows
linearly with time (as with $\epsilon=0$).
We have shown our results for $\epsilon =0.5$ in the inset in fig
\ref{fig:epdelta}.

Note that the above scaling analysis and our numerical simulation presented in fig
\ref{fig:eptag} and \ref{fig:epdelta} point towards $z_0/(2-\lambda) = 3$. If
$\lambda = b(\epsilon)$ as reported in~\cite{levine}, then for  
larger values of $\epsilon$ this would lead to $z_0$ smaller than
unity.    
For example, for $\epsilon =0.5$, we have verified that the above scaling form
remains valid (see  fig \ref{fig:epdelta}), which would imply $z_0 \simeq
0.54$ if $\lambda = b(\epsilon)$.

The other (simpler)
alternative is that $z_0=z=3/2$ and $\lambda = 3/2$ as in $\epsilon
=0$ case. This scenario would explain the observed $\rho_0$ dependence of
crossover time $\tau$. In the case of two probes, the above value of
$\lambda$ is 
consistent with the large time growth exponent of the average separation
$\langle r(t) \rangle$ between the probe pair (shown in fig \ref{fig:2zrav})
and also with the linear growth of $\Delta (t)$ shown in the inset of 
fig \ref{fig:epdelta}. 

\section{Conclusion}

In this paper, we have studied the dynamics of shock-tracking probe
particles in a one-dimensional KLS model of driven particles with
nearest neighbor Ising interaction $\epsilon$. In particular, we have
examined our results in the light of two different theoretical
scenarios.  The first scenario is based on an approximate mapping of
the problem to a zero-range process, and leads to the conclusion that
critical exponents characterizing power-law decays depend continuously
on the strength of interaction, $\epsilon$.  The second scenario is
based on the premise that since the correlations induced by Ising
interactions are short-ranged, asymptotic scaling properties which
involve large distances and large times 
should be independent of $\epsilon$, and the same
as at $\epsilon = 0$.  The results of our numerical studies on
the dynamical properties of the probe particles lend support to the second
scenario.

We find that in presence of only two probe particles in the system,
starting from a steady state configuration where the two probes were
nearest neighbors, the average distance $\langle r(t)\rangle$ between
them shows a crossover in time. For an initial time regime $\langle
r(t) \rangle$ the growth is consistent with a power law with an
$\epsilon$-dependent 
exponent $[2-b(\epsilon)]/z$~\cite{levine,rakos}.  However, for large
enough time, the growth occurs with an exponent $\simeq 1/3$, the
value expected for a noninteracting medium, consistent with the second
scenario discussed above. In addition, our study of the cumulative
distribution of the probe-separation shows that for large time, the
distribution function does not follow an $\epsilon$-dependent scaling form as
claimed in~\cite{rakos} but can be described by a form expected for
$\epsilon=0$, which again supports the second scenario mentioned above.

For a small but finite density of the probes, the dynamical
correlation functions show a similar scaling form as for $\epsilon
=0$~\cite{sc}. These scaling forms involve a crossover time-scale
$\tau$ that diverges for small $\rho_0$ as $\rho_0 ^{-3}$, as found
for the non-interacting case~\cite{sc}.  We have seen that an
$\epsilon$-dependent exponent $b(\epsilon)$ would lead to an
$\epsilon$-dependent dynamical exponent $z_0$ which may even become
less than unity for larger values of $\epsilon$. The other option, 
an $\epsilon$-independent dynamical exponent $z_0=3/2$, is consistent
with the second scenario 
outlined above, according to which, turning on a short-ranged Ising
interaction in the medium does not change the large time and large
distance properties of the system.

It is not yet completely
clear why the KLS-ZRP mapping does not seem to yield results which
agree with the numerical results. One possible reason is that the ZRP
mapping does not take into account the non-Markovian movement of the probes.
This lack of agreement also opens up the question of the nature of the
complete phase diagram for the problem under study, including negative values
of $\epsilon$?

\section*{Acknowledgements}
 
We acknowledge useful discussions with G.M. Sch\"{u}tz, D. Mukamel, Y. Kafri,
A. Rakos, M.R. Evans and D. Dhar. 


\end{document}